\documentclass[sn-nature]{sn-jnl}
\usepackage[table,cmyk]{xcolor}
\usepackage{amsmath,amsthm,,amsfonts,amssymb,commath,braket,booktabs}
\usepackage{calc,graphicx,bm,tikz,float}
\usepackage{xr}
\usepackage{multirow}
\usepackage{hyperref}
\definecolor{oeawblue}{cmyk}{0.9,0.68,0,0}
\definecolor{iqoqiblue}{cmyk}{0.76,0.11,0,0}
\definecolor{iffsred}{cmyk}{0.12,0.94,0.87,0.34}
\definecolor{thpurple}{cmyk}{0.65,1.0,0.0,0.2}
\definecolor{uestcblue}{cmyk}{0.99,0.78,0.16,0.03}

\usepackage{dcolumn}
\usepackage{braket}
\usepackage{xcolor}
\usepackage{tabularx}
\usepackage{mathtools}
\usepackage{pifont}
\usepackage{subcaption}
\usepackage{zhlipsum}
\usepackage{relsize}
\usepackage{subcaption}
\usepackage[numbers, sort&compress]{natbib}
\usepackage{soul} 

\usepackage{caption}
\usepackage{enumitem}
\setlist[enumerate]{after=\vspace{\baselineskip}}

\newcommand\blamb{\bm{\lambda}}
\newcommand\tr{{\rm tr}}
\newcommand{\comment}[1]{{}}

\hypersetup{
  pdftitle={Quantum Metrology under Indistinguishable Noise},
  pdfauthor={Lin et al.},
  pdfstartview=Fit,
  pdfpagelayout=SinglePage,
  colorlinks,
  linkcolor=uestcblue,
  citecolor=uestcblue,
  urlcolor=thpurple}
\usepackage[figure]{hypcap}

\usepackage[percent]{overpic}

\begin{document}

\title{Complementing Quantum Error Correction in Quantum Metrology via Swap Test}

\author[1,4]{\fnm{Xiaodie} \sur{Lin}}
\author[1]{\fnm{Linxuan} \sur{Li}}
\author*[1,2,3]{\fnm{Haidong} \sur{Yuan}}\email{hdyuan@mae.cuhk.edu.hk}

\affil[1]{\orgdiv{Department of Mechanical and Automation Engineering}, \orgname{The Chinese University of Hong Kong}, \orgaddress{\city{Hong Kong SAR}, \country{China}}}

\affil[2]{\orgdiv{The Hong Kong Institute of Quantum Information Science and Technology}, \orgname{The Chinese University of Hong Kong}, \orgaddress{\city{Hong Kong SAR}, \country{China}}}

\affil[3]{\orgdiv{State Key Laboratory of Quantum Information Technologies and Materials}, \orgname{The Chinese University of Hong Kong}, \orgaddress{\city{Hong Kong SAR}, \country{China}}}

\affil[4]{\orgdiv{College of Computer and Data Science}, \orgname{ Fuzhou University}, \orgaddress{\city{Fuzhou}, \postcode{350116}, \country{China}}}

\abstract{
The precision and sensitivity achievable in quantum metrology are often compromised by the presence of noise. While quantum error correction has emerged as a promising strategy, it is ineffective in addressing noise that is indistinguishable from the signal. To address this challenge, virtual state purification was introduced as a complementary approach to quantum error correction. However, significant noise accumulation can impede its performance. To overcome this limitation, we propose a swap test-based method specifically designed to address indistinguishable noise, even under high noise levels. A systematic error-scaling analysis demonstrates that this method enables quantum-enhanced precision in certain scenarios. Furthermore, numerical simulations demonstrate that our method surpasses virtual state purification in both single- and multi-parameter estimation tasks. The significant improvements in precision across diverse settings underscore the robustness and practicality of our method for real-world applications.
}

\maketitle

\section{Introduction}

Quantum metrology is a promising application of quantum science that uses entanglement, superposition, and other quantum features to enable measurements with sensitivity beyond classical limits. It has wide applications across diverse fields such as navigation~\cite{Feng_2019,nawrocki2015introduction}, communication~\cite{kohlrus17,conlon24}, and imaging~\cite{RevModPhys.71.1539,berchera2019quantum}. However, the potential advantages of quantum metrology are often compromised by decoherence and other noise sources, which can severely degrade achievable precision ~\cite{Davidovich2011,Madalin2012,Dobrzanski18,Fujiwara_2008,https://doi.org/10.1002/qute.202100080}.

Quantum error correction (QEC) has emerged as a promising strategy to counteract noise in quantum metrology, with many studies demonstrating its potential to preserve or enhance measurement precision~\cite{PhysRevLett.112.150802,PhysRevLett.112.080801,robust2015,PhysRevLett.115.200501,PhysRevLett.122.040502,Liang2018,PhysRevX.7.041009,PRXQuantum.2.010343}. A key limitation, however, is established by the \emph{Hamiltonian-not-in-Lindblad-span (HNLS)} condition: if the signal Hamiltonian $H_{\bm{\lambda}}$ lies within the span of the Lindblad operators, even infinite rounds of error correction cannot surpass the standard quantum limit~\cite{Liang2018,PhysRevX.7.041009}. In this scenario, the noise is indistinguishable from the signal, preventing the design of effective QEC codes and rendering Heisenberg-limited enhancement unattainable even with ancilla-assisted correction.


To address this challenge, QEM is introduced as a complementary approach to QEC in quantum metrology~\cite{kwon2025virtualpurificationcomplementsquantum}. Quantum error mitigation (QEM) offers an alternative approach by reducing noise-induced bias through classical post-processing of outputs from an ensemble of noisy quantum circuits~\cite{PhysRevLett.129.250503,PhysRevA.109.022410,hama2023quantum}. Unlike QEC, QEM does not correct errors within individual circuits but instead mitigates their effect statistically~\cite{RevModPhys.95.045005}. Specifically, one hybrid strategy uses a stabilizer state as the probe. QEC corrects the noise components orthogonal to the signal Hamiltonian \(H_{\boldsymbol{\lambda}}\), while the QEM technique of virtual state purification (VSP)~\cite{PhysRevX.11.041036,PhysRevX.11.031057}, mitigates the bias from the remaining noise components that lie within the span of \(H_{\boldsymbol{\lambda}}\). Although QEM inherently trades off an increase in variance, its targeted application—only correcting indistinguishable noise—results in a smaller variance overhead than protocols relying solely on QEM. Thus, the integration of QEC and QEM is synergistic and mutually beneficial.

However, VSP is effective primarily when the noiseless component remains the dominant part of the noisy output state. In quantum metrology, repeated use of a quantum channel for parameter encoding can lead to significant noise accumulation, which can diminish this dominance and thereby compromise VSP's effectiveness. To overcome this limitation, we propose a quantum metrology protocol based on the swap test ~\cite{doi:10.1137/S0097539796302452}. This method directly estimates error rates in order to construct a (nearly) unbiased estimator for the unknown parameters. We provide an analytical error-scaling analysis of the protocol, demonstrating that it can achieve quantum-enhanced precision across various scenarios. Furthermore, we numerically benchmark its performance against existing methods, showing substantial improvements in estimation precision. These results highlight the practical potential and robustness of our approach for real-world quantum sensing applications.

\section{Preliminaries}

\begin{figure}[htbp]
  \captionsetup[subfigure]{justification=centering, labelfont=bf}
  \centering
  \begin{subfigure}[t]{0.65\textwidth}
    \centering
    \includegraphics[width=\textwidth]{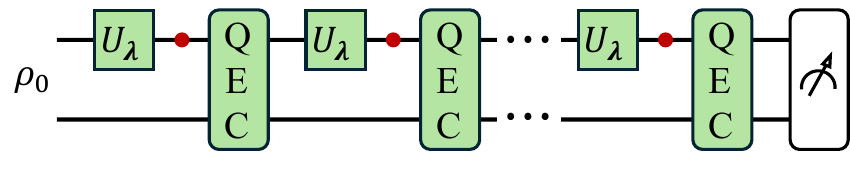}
  \end{subfigure}%
  \caption{\textbf{Scheme of quantum metrology.} A probe state $\rho_0$ is prepared and then subjected to $t/\Delta t$ rounds of iterative evolution. In each round, the state evolves under the noisy encoding unitary $U_{\bm{\lambda}}$, where the red dot represents the noise associated with $U_{\bm{\lambda}}$. Then, quantum error correction (QEC) is applied to eliminate correctable errors. Following the final round, information about $\bm{\lambda}$ is extracted by measuring on the final state.}\label{fig:setting}
\end{figure}

In a typical quantum metrology protocol, an unknown set of parameters \(\bm{\lambda}=(\lambda_1,\lambda_2,\cdots,\lambda_N)\) is encoded into a quantum state $\rho_{\bm{\lambda}}=e^{-itH_{\bm{\lambda}}}\rho_0U_{\bm{\lambda}}e^{itH_{\bm{\lambda}}}$ via the unitary evolution governed by the Hamiltonian $H_{\bm{\lambda}}$. Information about \(\bm{\lambda}\) is then extracted by performing a positive operator-valued measurement (POVM), \(\{E_x\}\) with \(\sum_x E_x = I\). The Born rule gives the outcome probability distribution \(P(x|\bm{\lambda}) = \operatorname{tr}(E_x \rho_{\bm{\lambda}})\). Repeating the procedure yields a sequence of measurement outcomes, from which an estimator \(\hat{\bm{\lambda}}=(\hat{\lambda}_1,\hat{\lambda}_2,\cdots,\hat{\lambda}_N)\) can be constructed. To protect such a procedure from noise, QEC can be applied to effectively eliminate correctable errors. Figure~\ref{fig:setting} illustrates this scheme, where state preparation, measurement and QEC operators are assumed to be noiseless, and noise (red dot) is introduced during the parameter-dependent evolution. In practice, state preparation and measurement can be implemented fault-tolerantly where applicable. Alternatively, high-fidelity probe state preparation can be accomplished using techniques such as entanglement distillation~\cite{PhysRevLett.76.722,PhysRevLett.77.2818} or similar distillation methods~\cite{zhao2021practical,liu2025dynamiclocccircuitsautomated}.

Unfortunately, when the signal  Hamiltonian $H_{\blamb}$ is within the Lindblad span, the noise becomes indistinguishable from the signal, rendering quantum error correction (QEC) ineffective~\cite{PhysRevX.7.041009,Liang2018,PRXQuantum.2.010343}.  To address this limitation, VSP~\cite{PhysRevX.11.041036,PhysRevX.11.031057} has been introduced to complement QEC for error mitigation~\cite{kwon2025virtualpurificationcomplementsquantum}. The core idea of VSP is to employ $m$ copies of the target noisy state  $\rho$ to measure the expectation values corresponding to the virtual state 
\begin{equation*}
    \bar{\rho}^m \coloneqq \frac{\rho^m}{\tr(\rho^m)}=\frac{\sum_i p_i^m\ket{i}\bra{i}}{\sum_ip_i^m},
\end{equation*}
where $\rho=\sum_i p_i\ket{i}\bra{i}$ is the spectral decomposition of $\rho$. When the noiseless state remains the dominant component in the evolved noisy state, VSP can effectively suppress indistinguishable noise. 

Concretely, consider a signal Hamiltonian defined as $H_{\lambda}=\lambda\sum_i Z^{(i)}$ and independent, identically distributed Pauli (IIDP) noise of the form
\begin{equation}\label{eq:iidp}
    \mathcal{E}^{(i)}(\rho)=p_I\rho+p_xX^{(i)}\rho X^{(i)}+p_yY^{(i)}\rho Y^{(i)}+p_zZ^{(i)}\rho Z^{(i)},
\end{equation}
where $X^{(i)},Y^{(i)},Z^{(i)}$ are Pauli operators acting on the $i$-th qubit. Then, the indistinguishable noise can be mitigated as follows~\cite{kwon2025virtualpurificationcomplementsquantum}: (1) Prepare a stabilizer probe state $\ket{\psi_0}$, stabilized by a group $G_s$, such that $X^{(i)},Y^{(i)},Z^{(i)}\notin G_s$; (2) Encode via the unitary $U_{\blamb}=e^{-itH_{\blamb}}$ to obtain the ideal logical output state $\ket{\psi}\coloneqq U_{\lambda}\ket{\psi_0}$; (3) Define the commuting stabilizer group $G_{s^{[c]}}\coloneqq \{S|[S,H_{\lambda}]=0,S\in G_s\}$, which ensures that $U_{\blamb}$ serves as a logical operation without introducing noise; (4) (optional) measure the noisy output state using the stabilizers in $G_{s^{[c]}}$ and correct Pauli errors in $X$ direction; (5) apply VSP to suppress noisy components in the output state that are orthogonal to $\ket{\psi}$, including $X^{(i)}\ket{\psi},Y^{(i)}\ket{\psi}$ and $Z^{(i)}\ket{\psi}$. Hence, such a VSP-based method can suppress single-qubit Pauli errors no matter whether (4) is executed or not.

Notice that the noisy state produced by a single Pauli $Z$ error---i.e., $Z^{(i)}\ket{\psi}$---remains within the logical space of the quantum error-correcting code. Consequently, QEC cannot detect or correct such an error. In contrast, VSP can effectively suppress errors of this type, leading to a significant performance advantage over QEC alone. Specifically, under IIDP noise with $p_I=O(1),p_x,p_y=\Theta(\Delta)$ and $p_z=O(\Delta)$ or $p_I=O(1),p_x,p_y=O(\Delta)$ and $p_z=\Theta(\Delta)$, the bias of QEC estimate scales as $O(\Delta)$. In comparison, VSP can achieve a bias that scales as $O(\Delta^m)$ for $m\ge 2$~\cite{kwon2025virtualpurificationcomplementsquantum}.

\section{Method}

As a concrete example, consider an $n$-qubit signal Hamiltonian $H_{\lambda}=\lambda\sum_iZ^{(i)}$ under local dephasing noise, described by the channel $\mathcal{E}^{(i)}_z(\rho)=(1-p_z)\rho+p_zZ^{(i)}\rho Z^{(i)}$. According to the Hamiltonian Not-in-Lindblad-Span (HNLS) condition, QEC cannot correct such a noise model. Consequently, the probe state $\ket{\psi_0}$ must undergo the noisy signal encoding, producing the noisy output state $\tilde{\rho}_z=\circ_i\mathcal{E}_{z}^{(i)}(\ket{\psi}\bra{\psi})$, where $\ket{\psi}=U_{\lambda}\ket{\psi_0}=e^{-itH_{\lambda}}\ket{\psi_0}$ is the ideal noiseless state.
    
To mitigate this noise, we adapt a strategy from VSP-based quantum metrology. Specifically, we choose the stabilizer probe state $|\psi_0\rangle=\frac{1}{\sqrt{2}}\left(\ket{0}^{\otimes n}+\ket{1}^{\otimes n}\right)$, which is stabilized by the stabilizer group $G_s=\langle X^{(1)}X^{(2)}\cdots X^{(n)},Z^{(1)}Z^{(2)},Z^{(2)}Z^{(3)},\cdots,Z^{(n-1)}Z^{(n)}\rangle$. Since  $\{X^{(1)}X^{(2)}\cdots X^{(n)},Z^{(i)}\}= 0$, directly using $G_s$ would misinterpret the signal as errors. Therefore, we remove the generator $X^{(1)}X^{(2)}\cdots X^{(n)}$ and define the new stabilizer group $G_{s^{[c]}}=\langle Z^{(1)}Z^{(2)},Z^{(2)}Z^{(3)},\cdots,Z^{(n-1)}Z^{(n)}\rangle$. Unlike standard VSP-based methods, that reduces the bias to $O(p_z^m)$ using $m$ copies of $\tilde{\rho}_z$~\cite{kwon2025virtualpurificationcomplementsquantum}, our goal here is to construct an unbiased estimator directly using only two copies of $\tilde{\rho}_z$.

\begin{figure}[t]
    \captionsetup[subfigure]{justification=centering, labelfont=bf}
    \centering
    \begin{subfigure}[t]{0.35\textwidth}
    \centering
    \includegraphics[width=\textwidth]{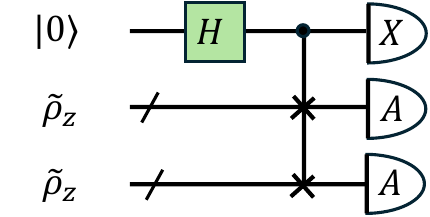}
    \end{subfigure}%
    \caption{\textbf{Circuit for the swap test.} By inputting two copies of the quantum state $\tilde{\rho}_z$, the expectation value $\langle X^{(1)} \rangle$ equals to $\text{tr}(\tilde{\rho}_z^2)$. Furthermore, the observable $A$ can be independently applied to each copy to obtain the expectation value $\tr(A\tilde{\rho}_z)$.}\label{fig:swap_test}
\end{figure}
    
The noisy output state subjected to $\mathcal{E}^{(i)}_z$ can be written as
\begin{equation}\nonumber
    \tilde{\rho}_z=(1-p_{\rm odd})\ket{\psi}\bra{\psi}+p_{\rm odd}\ket{\psi^{\perp}}\bra{\psi^{\perp}},
\end{equation}
where $p_{\rm odd}=\mathlarger{\sum}_{i{\rm\ is\ odd}}\begin{pmatrix}
    n \\
    i
\end{pmatrix}p_z^i(1-p_z)^{n-i}=\frac{1}{2}\left(1-(1-2p_z)^n\right)$, and $\ket{\psi^{\perp}}=\frac{1}{\sqrt{2}}U_{\lambda}\left(\ket{0}^{\otimes n}-\ket{1}^{\otimes n}\right)$. 
This indicates that the output states corresponding to even and odd Pauli $Z^{(i)}$ errors are orthogonal to each other. By applying the swap test on two copies of $\tilde{\rho}_z$, the overlap between them can be determined~\cite{doi:10.1137/S0097539796302452}. The corresponding circuit is depicted in Fig.~\ref{fig:swap_test}, where the expectation value of the Pauli $X$ operator acting on the control qubit is given by
\begin{equation}\label{eq:tr2_qec}
    \langle X^{(1)}\rangle = \tr(\tilde{\rho}_z^2)=1-2p_{\rm odd}+2p^2_{\rm odd}.
\end{equation}
Assuming a practical local error rate $p_z < \frac{1}{2}$, the probability $p_{\text{odd}}$ can be evaluated as $p_{\rm odd}=\frac{1-\sqrt{2\langle X^{(1)}\rangle-1}}{2}$. Furthermore, the target observable $A$ can be directly applied to these two copies of $\tilde{\rho}_z$  to evaluate the noisy expectation value $\tr(A\tilde{\rho}_z)$. Thus, both the error rate $p_{\rm odd}$ and the noisy expectation value $\tr(A\tilde{\rho}_z)$ can be simultaneously estimated using two copies of $\tilde{\rho}_z$. Specifically, the estimator of $\tr(A\tilde{\rho}_z)$ can be constructed as $\hat{A}=\frac{1}{2}(\hat{A}_2+\hat{A}_3)$, where $\hat{A}_i$ denotes the estimator of $\langle A^{(i)}\rangle$.

Based on such information, an analytic formulation can be derived to obtain an unbiased estimator for the interested parameter $\lambda$. For instance, let the observable $A=2\ket{\rm GHZ_y}\bra{\rm GHZ_y}-I$, where $\ket{\rm GHZ_y}=\frac{1}{\sqrt{2}}\left(\ket{0}^{\otimes n}-i\ket{1}^{\otimes n}\right)$. It follows that the unbiased estimator for $\lambda$ can be expressed as
\begin{equation}\label{eq:lambda_est}
    \hat{\lambda}=\arcsin\left(\frac{\hat{A}}{1-2\hat{p}_{\rm odd}}\right)/(2nt)=\arcsin\left(\frac{\hat{A}}{\sqrt{2\hat{X}-1}}\right)/(2nt),
\end{equation}
where $\hat{X}$ denotes the estimator of $\langle X^{(1)}\rangle$. 

Additionally, in the presence of IIDP noise, as described in Eq.~\eqref{eq:iidp}, the stabilizer group $G_{s^{[c]}}$ is capable of detecting both $X^{(i)}$ and $Y^{(i)}$ errors. However, these two error types yield identical syndromes for a given qubit $i$. If we interpret all such detected errors as $X^{(i)}$ errors and subsequently apply corrections, the noisy states $X^{(i)}\ket{\psi}\bra{\psi}X^{(i)}$ and $Y^{(i)}\ket{\psi}\bra{\psi}Y^{(i)}$ are transformed into $\ket{\psi}\bra{\psi}$ and $Z^{(i)}\ket{\psi}\bra{\psi}Z^{(i)}$, respectively. Consequently, when QEC is successfully applied, the IIDP noise is reduced to dephasing noise. This indicates that the previously described method remains applicable in this more generalized scenario. Although too many errors occurring simultaneously can lead to QEC failure, resulting in a biased estimator $\hat{\lambda}$, this bias converges to zero for a small local error rate as the number of qubits $n$ increases~\cite{nielsen2010quantum}. Moreover, our method can be extended to cases where orthogonality between noiseless and noisy states is not maintained. Please refer to~\ref{appdix:multiparas} for more discussion.

\section{Error analysis}

In quantum metrology, a primary objective is to achieve quantum-enhanced precision and sensitivity. The standard quantum limit (SQL) specifies an error scaling that is inversely proportional to either the evolution time $t$ or the number of qubits $n$, typically expressed as $1/t$ or $1/n$, respectively. Achieving a more rapid decrease in error scaling with $t$ or $n$ signifies a quantum advantage. However, in practice, longer evolution times or larger system sizes inevitably lead to greater noise accumulation. While QEC preserves quantum-enhanced precision in certain scenarios~\cite{PhysRevLett.112.150802,PhysRevLett.112.080801,robust2015,PhysRevLett.115.200501,PhysRevLett.122.040502,Liang2018}, it can fail when dealing with indistinguishable noise. In such cases, alternative mitigation techniques are required, but they can lead to an exponentially increased variance~\cite{PhysRevLett.131.210601,PhysRevLett.131.210602,endo22}. Therefore, it is meaningful to characterize the error scaling of the proposed method and observe whether quantum advantage exists in the presence of noise.

\subsection{Dephasing noise}
In the swap test-based method, the estimator $\hat{\lambda}$ defined in Eq.~\eqref{eq:lambda_est} relies on the estimators $\hat{A}$ and $\hat{X}$. Using the first-order Taylor expansion around $\langle A\rangle_{\tilde{\rho}_z}$ and $\langle X^{(1)}\rangle$, $\hat{\lambda}$ can be approximated as
\begin{equation}
    \hat{\lambda}\approx \lambda+\left(\frac{\partial \hat{\lambda}}{\partial\hat{A}}\right)\left(\hat{A}-\langle A\rangle_{\tilde{\rho}_z}\right)+\left(\frac{\partial \hat{\lambda}}{\partial\hat{X}}\right)\left(\hat{X}-\langle X^{(1)}\rangle\right).
\end{equation}
Then, it follows that
\begin{equation}\label{eq:var_lamb}
    {\rm Var}(\hat{\lambda})\approx \left(\frac{\partial \hat{\lambda}}{\partial\hat{A}}\right)^2{\rm Var}(\hat{A}) + \left(\frac{\partial \hat{\lambda}}{\partial \hat{X}}\right)^2{\rm Var}(\hat{X}) + 2\left(\frac{\partial \hat{\lambda}}{\partial\hat{A}}\right)\left(\frac{\partial \hat{\lambda}}{\partial\hat{X}}\right){\rm Cov}(\hat{A}, \hat{X}).
\end{equation}
Suppose the circuit illustrated in Fig.~\ref{fig:swap_test} is executed $\nu$ times, and the quantum state immediately before measurement is $\sigma$. Then, it holds that
\begin{equation}\label{eq:sigma}
    \sigma=\frac{1}{2}\left((\ket{0}\bra{0}+\ket{1}\bra{1})\otimes\tilde{\rho}_z^{\otimes2}+(\ket{0}\bra{1}+\ket{1}\bra{0})\otimes{\rm SWAP}\tilde{\rho}_z^{\otimes2}\right),
\end{equation}
since ${\rm SWAP}\tilde{\rho}_z^{\otimes2}=\tilde{\rho}_z^{\otimes2}{\rm SWAP}$. Now, we try to evaluate ${\rm Var}(\hat{\lambda})$. First, the variance of $\hat{A}$ reads ${\rm Var}(\hat{A})=\frac{1}{4}\left({\rm Var}(\hat{A}_2)+{\rm Var}(\hat{A}_3)+2{\rm Cov}(\hat{A}_2,\hat{A}_3)\right)$. According to Eq.~\eqref{eq:sigma}, it holds that ${\rm Cov}(\hat{A}_2,\hat{A}_3)=\langle A^{(2)}A^{(3)}\rangle-\langle A^{(2)}\rangle\langle A^{(3)}\rangle=0$. Meanwhile, since ${\rm Var}(\hat{A}_2)={\rm Var}(\hat{A}_3)\le \frac{1}{\nu}\Vert A^2\Vert_{\infty}=\frac{1}{\nu}$, it can be easily verified that ${\rm Var}(\hat{A})\le \frac{1}{2\nu}$ and
\begin{equation}\label{eq:var_a}
\begin{aligned}
    \left(\frac{\partial \hat{\lambda}}{\partial\hat{A}}\right)^2{\rm Var}(\hat{A})&=\left(4n^2t^2(2\langle X^{(1)}\rangle-1-\langle A\rangle_{\tilde{\rho}_z}^2)\right)^{-1}{\rm Var}(\hat{A})\\
    &\le \left(8\nu n^2t^2(2\langle X^{(1)}\rangle-1-\langle A\rangle_{\tilde{\rho}_z}^2)\right)^{-1}
\end{aligned},
\end{equation}
where $\Vert\cdot \Vert_{\infty}$ denotes the spectral norm. Second, note that ${\rm Var}(\hat{X})=\frac{1}{\nu}\left(\langle {X^{(1)}}^2\rangle-\langle X^{(1)}\rangle^2\right)=\frac{1}{\nu}\left(1-\langle X^{(1)}\rangle^2\right)\le \frac{3}{4\nu}$, since $\langle X^{(1)}\rangle=\frac{1}{2}+(1-2p)^n> \frac{1}{2}$ according to Eq.~\eqref{eq:tr2_qec}.
Similarly, we have
\begin{equation}\label{eq:var_x}
    \left(\frac{\partial \hat{\lambda}}{\partial \hat{X}}\right)^2{\rm Var}(\hat{X})\le 3\langle A\rangle_{\tilde{\rho}_z}^2\left(16\nu n^2t^2\left(2\langle X^{(1)}\rangle-1\right)^2\left(2\langle X^{(1)}\rangle-1-\langle A\rangle_{\tilde{\rho}_z}^2\right)\right)^{-1}.
\end{equation}
Last, since $\hat{A_2}$ and $\hat{A_3}$ are independent from each other, ${\rm Cov}(\hat{A},\hat{X})=\frac{1}{\nu}\left(\langle X^{(1)}A^{(2)}\rangle-\langle X^{(1)}\rangle\langle A^{(2)}\rangle\right)=\frac{1}{\nu}\left(\langle A\rangle_{\tilde{\rho}_z^2}-\langle X^{(1)}\rangle\langle A\rangle_{\tilde{\rho}_z}\right)$. Notice that $\langle A\rangle_{\tilde{\rho}_z^2}=\langle A\rangle_{\tilde{\rho}_z}$, hence $\langle A\rangle_{\tilde{\rho}_z}{\rm Cov}(\hat{A},\hat{X})=\frac{1}{\nu}\langle A\rangle_{\tilde{\rho}_z}^2\left(1-\langle X^{(1)}\rangle\right)\le \frac{1}{2\nu}\langle A\rangle_{\tilde{\rho}_z}^2$. Thus, it turns out that
\begin{equation}\label{eq:cov_ax}
\begin{aligned}
    &\left(\frac{\partial \hat{\lambda}}{\partial\hat{A}}\right)\left(\frac{\partial \hat{\lambda}}{\partial\hat{X}}\right){\rm Cov}(\hat{A}, \hat{X})\\
    =&\langle A\rangle_{\tilde{\rho}_z}{\rm Cov}(\hat{A},\hat{X})\left(4n^2t^2\left(2\langle X^{(1)}\rangle-1\right)\left(2\langle X^{(1)}\rangle-1-\langle A\rangle_{\tilde{\rho}_z}^2\right)\right)^{-1}\\
    \le&\langle A\rangle_{\tilde{\rho}_z}^2\left(8\nu n^2t^2\left(2\langle X^{(1)}\rangle-1\right)\left(2\langle X^{(1)}\rangle-1-\langle A\rangle_{\tilde{\rho}_z}^2\right)\right)^{-1}
\end{aligned}.
\end{equation}
Substituting Eqs.~\eqref{eq:var_a},~\eqref{eq:var_x}, and~\eqref{eq:cov_ax} into Eq.~\eqref{eq:var_lamb}, the final variance reads
\begin{equation}
\begin{aligned}
    {\rm Var}(\hat{\lambda})\le & \left(8\nu n^2t^2(1-2p_{\rm odd})^2\eta\right)^{-1}+3\sin^2(2nt\lambda)\left(16\nu n^2t^2(1-2p_{\rm odd})^4\eta\right)^{-1}\\
    & +\sin^2(2nt\lambda)\left(4\nu n^2t^2(1-2p_{\rm odd})^2\eta\right)^{-1}\\
    \le & 3\left(8\nu n^2t^2(1-2p_{\rm odd})^2\eta\right)^{-1}+3\left(16\nu n^2t^2(1-2p_{\rm odd})^4\eta\right)^{-1}\\
    = & \mathcal{O}\left(\left(\nu n^2t^2(1-2p_{\rm odd})^4\eta\right)^{-1}\right)
\end{aligned},
\end{equation}
where $\eta=1-\sin^2(2nt\lambda)$. The first inequality is derived using the relations $\langle A \rangle_{\tilde{\rho}_z}^2 = (1-2p_{\rm odd})^2 \sin^2(2nt\lambda)$ and $2\langle X^{(1)} \rangle - 1 = (1-2p_{\rm odd})^2$.
The term $(nt)^{-2}$ in the variance indicates the Heisenberg scaling with respect to both the number of qubits $n$ and the evolution time $t$. Additionally, the effect of noise is characterized by the factor $(1-2p_{\rm odd})^4$.

\begin{figure}[t]
    \captionsetup[subfigure]{justification=centering, labelfont=bf}
    \centering
    \begin{subfigure}[t]{0.6\textwidth}
    \centering
    \includegraphics[width=\textwidth]{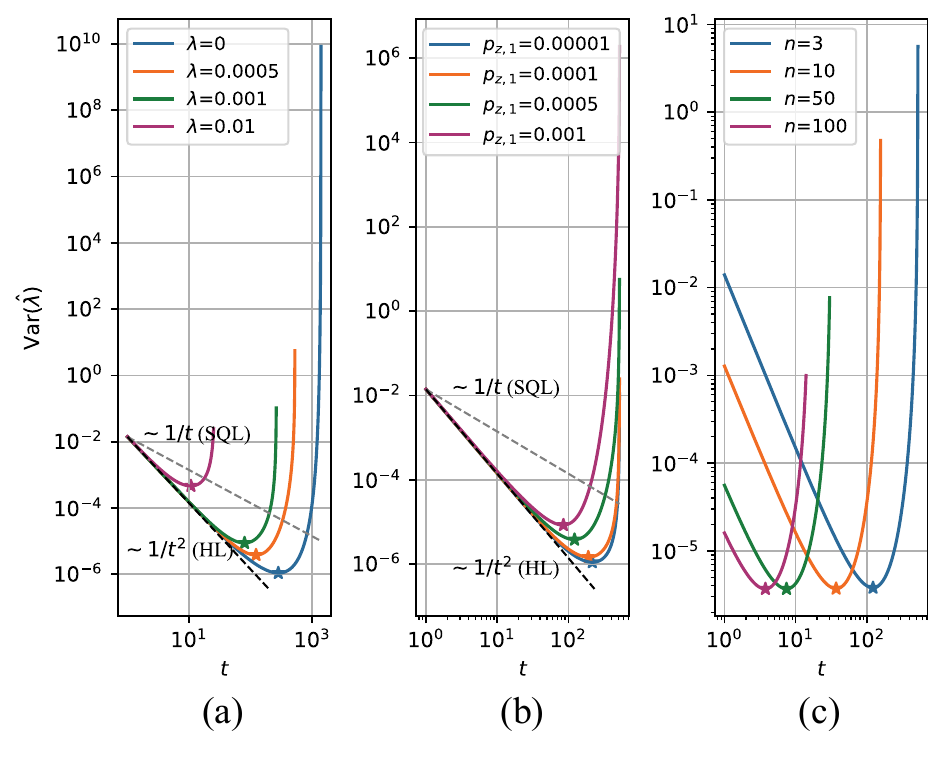}
    \end{subfigure}%
    \caption{\textbf{Scaling of the variance of the estimator $\hat{\lambda}$.} (a) Variance behavior for different values of the parameter $\lambda$, with an initial local error rate of $p_{z,1} = 0.0005$ and a qubit number of $n = 3$. (b) Variance behavior for varying values of $p_{z,1}$, where $\lambda = 0.0005$ and $n = 3$. (c) Variance behavior for different qubit numbers $n$, with fixed parameters $\lambda = p_{z,1} = 0.0005$. In each subplot, the minimal variance for each case is marked with a star. Additionally, the scaling for the standard quantum limit (SQL) and the Heisenberg limit (HL) are shown as gray and black dashed lines, respectively.}\label{fig:zeeman_var}
\end{figure}

Furthermore, consider the scenario where the local error rate accumulated after the evolution time $t$, denoted as $p_t$, follows the relation $p_{z,t} = 1 - (1-p_{z,1})^t$, where $p_{z,1}$ is the initial local error rate. Under this assumption, it can be shown that
\begin{equation}
1-2p_{\rm odd} = (1-2p_{z,t})^n = \left(2(1-p_{z,1})^t - 1\right)^n.
\end{equation}
Focusing on the evolution time $t$ such that $p_{z,t} < \frac{1}{2}$ with $\eta \neq 0$, Fig.~\ref{fig:zeeman_var} illustrates the scaling behavior of ${\rm Var}(\hat{\lambda})$. Specifically, Fig.~\ref{fig:zeeman_var}(a) presents the scaling for various values of $\lambda$ when the initial local error rate is set to $p_{z,1} = 0.0005$ and the number of qubits is $n=3$. An increase in $\lambda$ results in a larger variance for a fixed $t$. Moreover, as $\lambda$ increases, the overall minimal variance also becomes greater. Notably, smaller values of $\lambda$ exhibit a stronger quantum advantage, as reflected in the larger evolution time $t$ for which the variance remains within the region bounded by the gray and black dashed lines, representing the SQL and the Heisenberg limit (HL), respectively. This underscores the significant role of feedback control in enhancing the precision of parameter estimation.

As a comparison, the impact of varying $p_{z,t}$ on the variance is relatively minor, as depicted in Fig.~\ref{fig:zeeman_var}(b) for $\lambda = 0.0005$ and $n=3$. Additionally, the scaling behavior under varying qubit numbers $n$ is shown in Fig.~\ref{fig:zeeman_var} for the fixed parameters $\lambda = p_{z,1} = 0.0005$. When $t$ is sufficiently small, the variance exhibits Heisenberg scaling with respect to $n$. Interestingly, for different values of $n$, the minimal variance remains nearly unchanged. This suggests that adopting a smaller $n$ in practical experiments can effectively reduce the complexity while still ensuring comparable precision, provided the system undergoes evolution for a sufficiently long time.

\subsection{IIDP noise}

For simplicity, assume that the IIDP noise takes the form $\mathcal{E}^{(i)}=\mathcal{E}_z^{(i)}\circ\mathcal{E}_x^{(i)}$, where $\mathcal{E}^{(i)}_x(\rho)=(1-p_x)\rho+p_xX^{(i)}\rho X^{(i)}$. The analysis of error scaling can be extended to more general noise models using an analogous approach. Note that the $n$-qubit probe state can detect and correct Pauli $X$ errors when $X^{(i)}$ occurs on fewer than $\lceil n/2\rceil$ qubits. Otherwise, it would misinterpret errors, thus overall cause a global $X^{\otimes n}$ error to the state.
To enhance the success probability of QEC, we consider a protocol in which QEC is performed on the probe state after each time unit, with error rate $p_{x,1}$ and $p_{z,1}$ for the noise channels $\mathcal{E}_x^{(i)}$ and $\mathcal{E}_z^{(i)}$, respectively. Then, since the QEC operation is commute with $\mathcal{E}_z^{(i)}$, the noisy output state after evolution through $t$ time units under $\otimes_i \mathcal{E}^{(i)}$ is given by
\begin{equation}
    \tilde{\rho}=(1-p_{x,1}')^t\tilde{\rho}_z+\left(1-(1-p_{x,1}')^t\right)\tilde{\rho}_x,
\end{equation}
where $p_{x,1}'=\sum_{i=\lceil n/2\rceil}^{n}\begin{pmatrix} n \\ i \end{pmatrix}p_{x,1}^{i}(1-p_{x,1})^{n-i}$ denotes the the probability of bit-flip errors remaining after QEC application, and $\tilde{\rho}_x$ denotes the quantum state resulting from unsuccessful QEC operations.

According to Eq.~\eqref{eq:var_lamb}, the bias of the estimator $\hat{\lambda}$ can be approximated as
\begin{equation}\label{eq:bias_lamb_xz}
\begin{aligned}
    {\rm Bias}\left(\hat{\lambda}\right)=&\left|\mathbb{E}\left(\hat{\lambda}\right)-\lambda\right|\\
    \approx&\left|\left(\frac{\partial \hat{\lambda}}{\partial\hat{A}}\right)\left(\mathbb{E}\left(\hat{A}\right)-\langle A\rangle_{\tilde{\rho}_z}\right)+\left(\frac{\partial \hat{\lambda}}{\partial\hat{X}}\right)\left(\mathbb{E}\left(\hat{X}\right)-\langle X^{(1)}\rangle\right)\right|
\end{aligned},
\end{equation}
where $\mathbb{E}(\cdot)$ represents the expectation value of the estimator. Define $p''=(1-p_{x,1}')^t$, we have
\begin{equation}\label{eq:bias_a_xz}
\begin{aligned}
    \left|\mathbb{E}\left(\hat{A}\right)-\langle A\rangle_{\tilde{\rho}_z}\right|&=\left(1-p''\right)\left|\tilde{\rho}_x-\tilde{\rho}_z\right|\\
    &\le \left(1-p''\right)\Vert A\Vert_{\infty}\Vert \tilde{\rho}_x-\tilde{\rho}_z\Vert_1\\
    &\le 2\left(1-p''\right)
\end{aligned},
\end{equation}
and
\begin{equation}\label{eq:bias_x_xz}
\begin{aligned}
    \left|\mathbb{E}\left(\hat{X}\right)-\langle X^{(1)}\rangle\right|&=\left|-\left(1-p''^{2}\right)\langle X^{(1)}\rangle+\left(1-p''\right)^2\tr(\tilde{\rho}_x^2)+2p''\left(1-p'' \right)\tr(\tilde{\rho}_z\tilde{\rho}_x)\right|\\
    &\le 2\left(1-p''^2\right)
\end{aligned},
\end{equation}
since $\langle X^{(1)}\rangle,\tr(\tilde{\rho}_x^2),\tr(\tilde{\rho}_z\tilde{\rho}_x)\ge 0$. Consequently, the bias can be upper bounded by
\begin{equation}\label{eq:bias_lamb_xz_bound}
\begin{aligned}
    {\rm Bias}\left(\hat{\lambda}\right)&\le \left|\left(\frac{\partial \hat{\lambda}}{\partial\hat{A}}\right)\right|\cdot \left|\mathbb{E}\left(\hat{A}\right)-\langle A\rangle_{\tilde{\rho}_z}\right|+\left|\left(\frac{\partial \hat{\lambda}}{\partial\hat{X}}\right)\right|\cdot\left|\mathbb{E}\left(\hat{X}\right)-\langle X^{(1)}\rangle\right|\\
    &\le \left(1-p''\right)\left(nt\sqrt{2\langle X^{(1)}\rangle-1-\langle A\rangle_{\tilde{\rho}_z}^2}\right)^{-1}\\
    &\quad\ +\langle A\rangle_{\tilde{\rho}_z}\left(1-p''^2\right)\left(nt\left(2\langle X^{(1)}\rangle-1\right)\sqrt{2\langle X^{(1)}\rangle-1-\langle A\rangle_{\tilde{\rho}_z}^2}\right)^{-1}\\
    &=\frac{1-p''}{nt(1-2p_{\rm odd})\sqrt{\eta}}+\frac{\left(1-p''^2\right)\sin(2nt\lambda)}{nt(1-2p_{\rm odd})^2\sqrt{\eta}}\\
    &=\frac{\left(1-p''\right)}{nt(1-2p_{\rm odd})\sqrt{\eta}}\left(1+\frac{\left(1+p''\right)\sin(2nt\lambda)}{1-2p_{\rm odd}}\right)
\end{aligned}.
\end{equation}

\begin{figure}[t]
    \captionsetup[subfigure]{justification=centering, labelfont=bf}
    \centering
    \begin{subfigure}[t]{\textwidth}
    \centering
    \includegraphics[width=\textwidth]{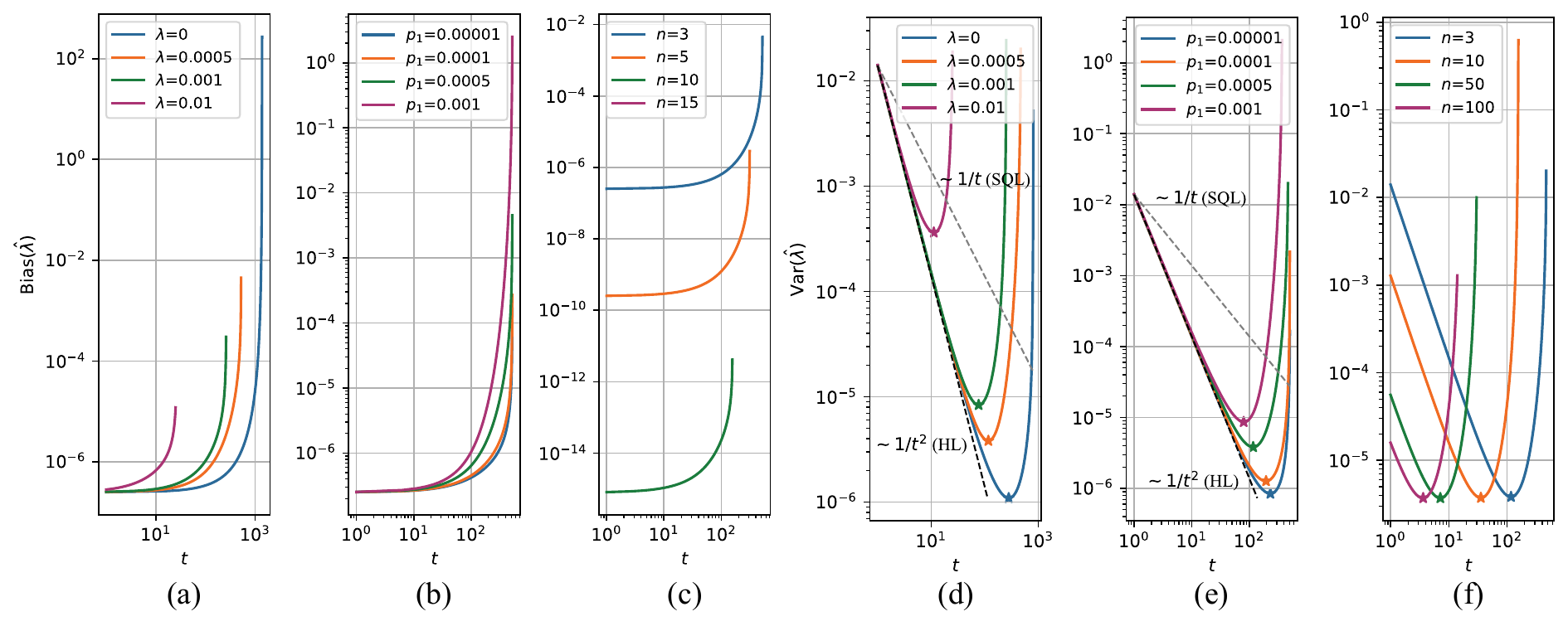}
    \end{subfigure}%
    \caption{\textbf{Error scaling of the estimator $\hat{\lambda}$ under IIDP noise.} Let $p_{z,1}=p_{x,1}=p_1$. \textbf{(a-c) Bias behavior:} Scaling of the bias for estimator $\hat{\lambda}$. Specifically, (a) shows the bias for various values of parameter $\lambda$, with a fixed initial local error rate $p_{1}= 0.0005$ and a qubit number of $n=3$. (b) illustrates the bias as a function of $p_{1}$, with $\lambda = 0.0005$ and $n=3$. (c) displays the bias for different qubit numbers $n$, with fixed parameters $\lambda = p_1 = 0.0005$.
    \textbf{(d-f) Variance behavior:} Scaling of the variance for estimator $\hat{\lambda}$. Specifically, (d) shows the variance for various values of parameter $\lambda$, with a fixed initial local error rate $p_1 = 0.0005$ and $n=3$. (e) illustrates the variance as a function of $p_{1}$, with $\lambda = 0.0005$ and $n=3$. (f) displays the variance for different qubit numbers $n$, with fixed parameters $\lambda = p_1 = 0.0005$. In each subplots, the minimal variance for each case is marked with a star. The dashed gray and black lines represent the scaling for the Standard Quantum Limit (SQL) and the Heisenberg Limit (HL), respectively.}\label{fig:zeeman_bias}
\end{figure}

Fig.~\ref{fig:zeeman_bias}(a-c) depicts the scaling of bias across different cases. First, note that ${\rm Bias}\left(\hat{\lambda}\right)$ increases with $\lambda$. Fig.~\ref{fig:zeeman_bias}(a) visualizes the corresponding biases for different values of $\lambda$ and demonstrates the increased bias caused by accumulated noise. Additionally, Fig.~\ref{fig:zeeman_bias}(b) compares behaviors under different local error rate levels. For small $t$, they vary slightly, while differing by several orders of magnitude when $2nt\lambda$ approaches $\frac{\pi}{2}$. Finally, Fig.~\ref{fig:zeeman_bias}(c) shows significant differences in bias for different logical state sizes, where increasing the qubit number $n$ from 3 to 10 leads to a bias decrease of approximately $10^8$. When $n=15$, the estimator becomes nearly unbiased.

Regarding the variance calculation, we can employ a first-order Taylor expansion around $\langle A\rangle_{\tilde{\rho}}$ and $\tr(\tilde{\rho}^2)$ to approximate $\hat{\lambda}$, subsequently applying an analogous analysis to the case where only dephasing noise is present. A detailed calculation is provided in \ref{appdix:variance_iidp}. The corresponding upper bound of ${\rm Var}(\hat{\lambda})$ is illustrated in Fig.~\ref{fig:zeeman}(d-f), where the curves are truncated when these bounds become positive as the evolution time $t$ increases. Compared to the case with solely dephasing noise, the presence of bit-flip noise has only a slight impact on the variance. Again, for fixed local error rates and $\lambda$, the minimum variance decreases with increasing $n$, while varying little.

\section{Numerical simulations}

To evaluate the performance of our swap test-based method, we investigate its behaviors in both single- and multi-parameter estimation tasks. In each scenario, we compared its performance against two other approaches: the original noisy method and a VSP-based method. For a fair comparison, we specifically focused on the second-order VSP-based method (i.e., with $m=2$). Our numerical results demonstrate the significant advantage of the swap test-based method in enhancing estimation precision, highlighting its promising applicability in quantum metrology to address challenges posed by indistinguishable noise.

\subsection{Single-parameter estimation}

For the canonical phase estimation task with the Hamiltonian $H_{\lambda}=\lambda\sum_iZ^{(i)}$, we initially investigate the performance of different methods under dephasing noise. Specifically, the system parameters were set as the number of qubits, $n=3$, and the unknown parameter, $\lambda=0.001$. Furthermore, the accumulated error rate, $p_{z,t}$, resulting from local dephasing noise after evolution time $t$, is modeled to increase according to $p_{z,t} = 1 - (1-p_{z,1})^t$, where  $p_{z,1} = 0.0005$.

\begin{figure}[t]
  \centering
  \includegraphics[width=\columnwidth]{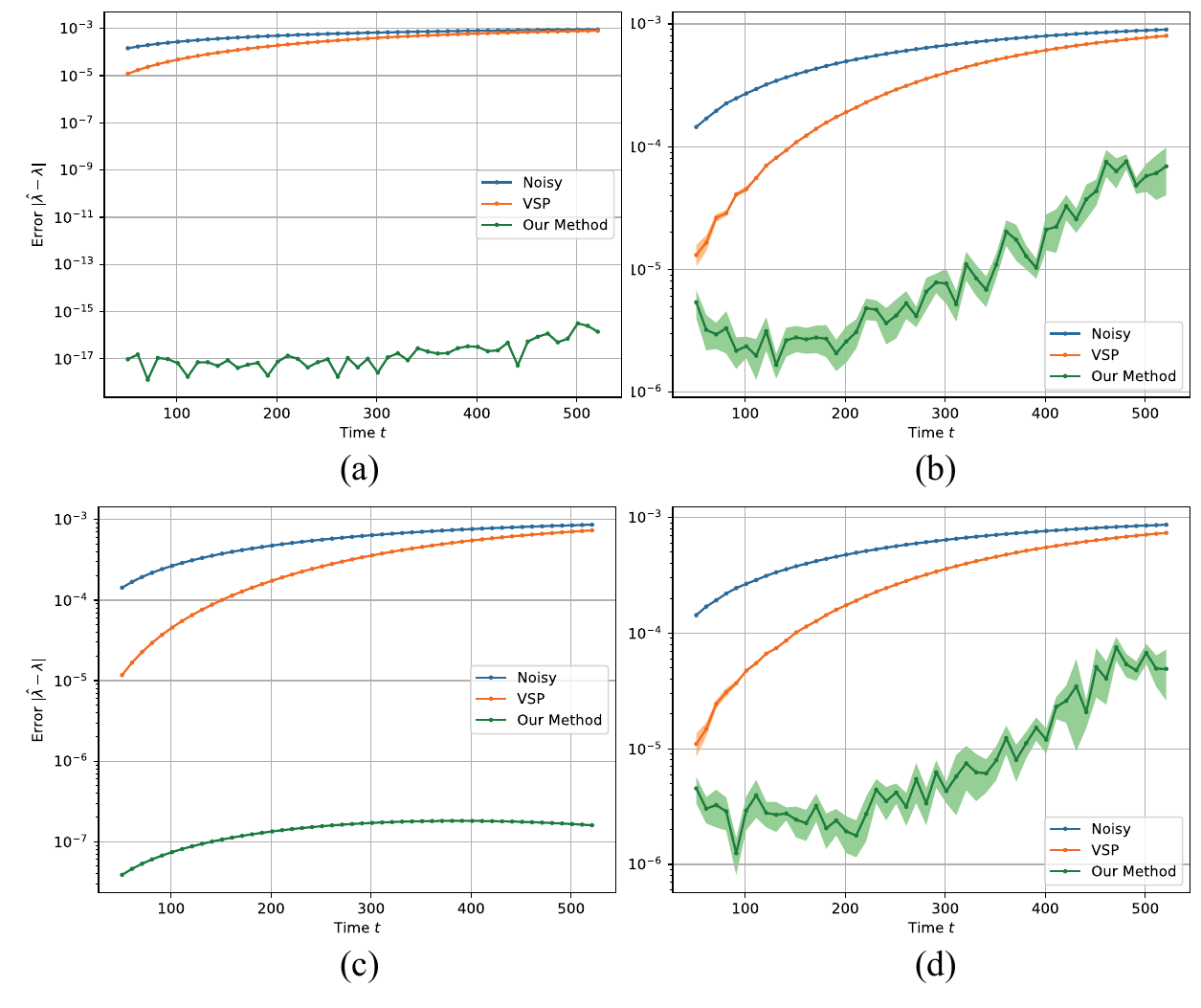}
  \caption{\textbf{Single-parameter estimation errors for different methods.} The absolute parameter estimation errors, $|\hat{\lambda} - \lambda|$, are shown for the original noisy (blue), VSP-based (orange), and our (green) methods. Panels (a) and (b) show results under dephasing noise, while (c) and (d) show results under IIDP noise. Performance is compared for infinite measurement shots ((a), (c)) and $10^6$ shots ((b), (d)). In particular, numerical simulations in panels (b) and (d) are performed 10 times to compute the mean absolute errors and the corresponding 95\% confidence intervals, represented by solid lines and shaded regions, respectively.}\label{fig:zeeman}
\end{figure}

Fig.~\ref{fig:zeeman}(a) and (b) illustrate the parameter estimation errors, $|\hat{\lambda}-\lambda|$, for the three methods, corresponding to infinite and $10^6$ measurement shots, respectively. These results were obtained by selecting evolution times ranging from 1 to $\lfloor \pi/(2n\lambda)\rfloor$ with an interval of 10.
In the case of infinite measurement shots, our method demonstrates unbiased estimation, with observed errors primarily caused by numerical precision limitations. Conversely, both the original noisy method and the VSP-based method exhibit errors ranging from $10^{-5}$ to $10^{-3}$. This significant advantage of our method is maintained even when a finite number of measurement shots is available.
Specifically, with $10^6$ measurement shots and after repeating the experiment 10 times, the mean parameter estimation errors for each method are represented by solid lines, while the shaded areas indicate the corresponding 95\% confidence intervals. As evidenced by these results, our method consistently achieves a significantly reduced error, on the order of magnitudes lower than other methods.

To further investigate the performance of the proposed approach in the presence of more complex noise, we consider the IIDP noise of the form $\mathcal{E}^{(i)}=\mathcal{E}_z^{(i)}\circ\mathcal{E}_x^{(i)}$. Setting the local error rates to $p_{z,1}=p_{x,1}=0.0005$ at each time step, QEC is implemented at every time unit to suppress detectable errors, effectively reducing the local $X^{(i)}$ error rate, $p_{x,1}$, to a global $X^{\otimes n}$ error rate of approximately $p_{x,1}'\approx 7.5\times10^{-7}$.

Maintaining the experimental setup described above, Fig.~\ref{fig:zeeman}(c) and (d) illustrate the corresponding results. Since the $X$-type error rate is suppressed to a relatively low value, the proposed method still achieves significantly reduced parameter estimation errors compared to both the original noisy and VSP-based methods. Moreover, when compared to the scenario involving only dephasing noise, the overall performance of all three methods is comparable to the preceding results. Notably, the comparison of the green line in Fig.~\ref{fig:zeeman}(c) with the orange curve in Fig.~\ref{fig:zeeman_bias}(a) highlights that the estimation bias of the proposed method remains below $10^{-6}$ over time, rather than increasing rapidly with the evolution time $t$. This finding underscores the robust and reliable performance of the proposed approach in practical applications.

\subsection{Multi-parameter estimation}\label{subsec:multi}

The swap test-based method can be naturally extended to the multi-parameter estimation tasks. For instance, consider a signal Hamiltonian $H_{\blamb}=\lambda_1X^{(1)}X^{(2)}\cdots X^{(n)}+\lambda_2\sum_{i=1}^nZ^{(i)}$, where $\blamb=(\lambda_1,\lambda_2)$. Again, set the probe state as $|\psi_0\rangle=\frac{1}{\sqrt{2}}\left(\ket{0}^{\otimes n}+\ket{1}^{\otimes n}\right)$, the stabilizer group $G_{s^{[c]}}$ remains $\langle Z^{(1)}Z^{(2)},Z^{(2)}Z^{(3)},\cdots,Z^{(n-1)}Z^{(n)}\rangle$. Hence, QEC can be implemented at each time unit to detect and correct bit-flip errors. However, since $[Z^{(i)},H_{\blamb}]\neq0$, the noisy evolved state $Z^{(i)}\ket{\psi}$ is not necessarily orthogonal to the noiseless state $\ket{\psi}$. Nevertheless, it can be formally demonstrated that for small $\lambda_i$, these states remain approximately orthogonal. Moreover, even for larger values of $\lambda_i$, our swap test-based approach can be naturally adjusted to accommodate such scenarios effectively. Further details can be found in~\ref{appdix:multiparas}.

\begin{figure}[t]
    \captionsetup[subfigure]{justification=centering, labelfont=bf}
    \centering
    \begin{subfigure}[t]{\textwidth}
    \centering
    \includegraphics[width=\textwidth]{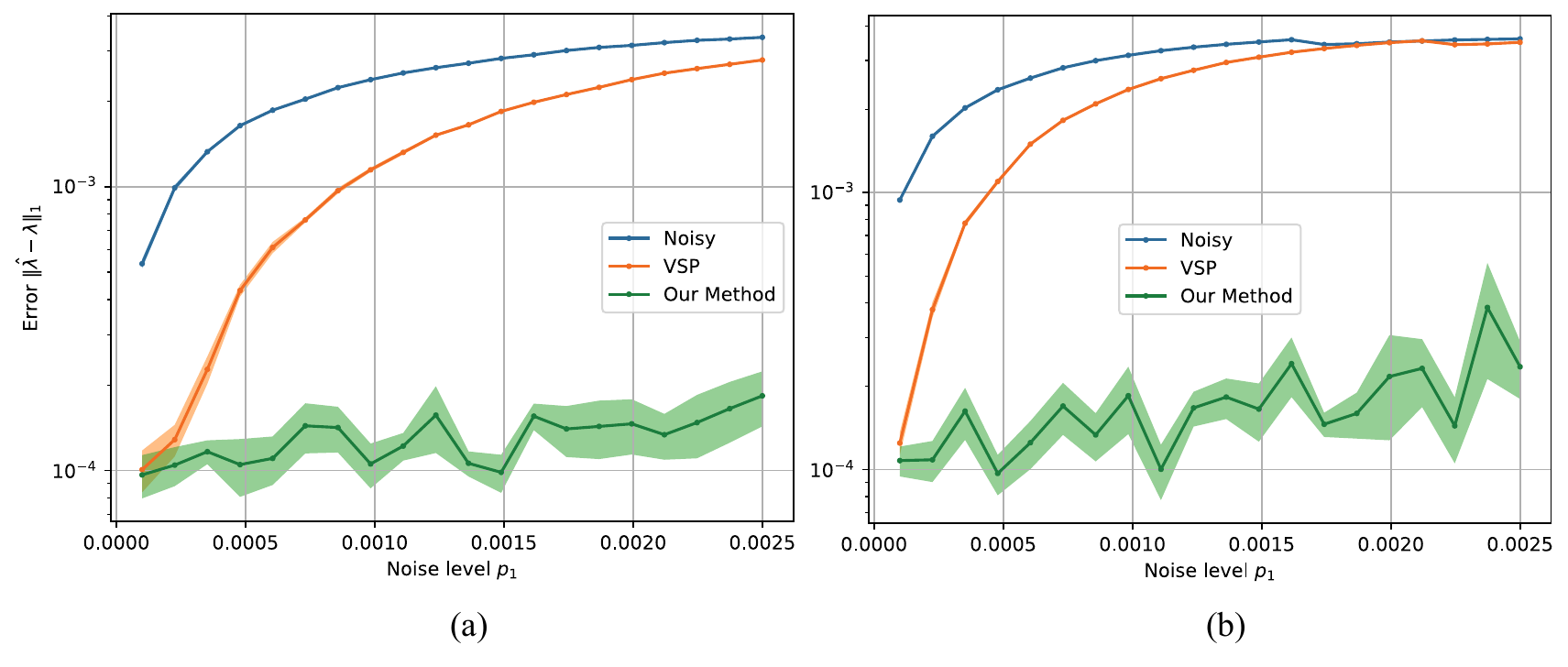}
    \end{subfigure}%
    \caption{\textbf{Multi-parameter estimation errors for different methods.} The absolute parameter estimation errors, $\Vert\hat{\blamb} - \blamb\Vert_1$, are shown for the original noisy (blue), VSP-based (orange), and our (green) methods. Panel (a) shows results under dephasing noise, while (b) shows results under IIDP noise. For each scenario, numerical simulations were performed 10 times with $10^6$ measurement shots to determine the mean absolute errors, indicated by solid lines, and their corresponding 95\% confidence intervals, shown as shaded regions.}\label{fig:multi}
\end{figure}

Let $\blamb = (0.001, 0.002)$, $n = 3$, and the total evolution time $t = 100$. By specifying the observable as $Y^{\otimes n}$ with a measurement shot number of $10^6$, the unknown parameters can be estimated using the maximum likelihood estimator (MLE)~\cite{380f7170-a649-307c-9495-f3b3298846ff}. To reduce the influence of local minima and focus solely on the degree of noise suppression achieved by each method, the parameters are randomly initialized near their true values before applying MLE to obtain the estimated results. Figs.~\ref{fig:multi}(a) and (b) depict the corresponding estimation errors, $\Vert\hat{\blamb} - \blamb\Vert_1$, of different methods under two noise models: dephasing noise with a local error rate $p_{z,1} = p_1$, and IIDP noise with local error rates $p_{x,1} = p_{y,1} = p_{z,1} = p_1$, respectively.
As the local error rate $p_1$ increases from $0.0001$ to $0.0025$, the performance of the VSP-based method deteriorates, converging to that of the original noisy method. In contrast, our swap test-based method demonstrates consistently low estimation errors under dephasing noise, although the variance increases slightly as $p_1$ rises. Under IIDP noise, however, the performance of our method degrades with increasing $p_1$, primarily due to the decreasing success probability of QEC. Specifically, the probability of QEC misinterpreting syndromes per time unit increases from $1.20 \times 10^{-7}$ to $7.48 \times 10^{-5}$, leading to a significant accumulation of bit-flip errors over 100 time units.
Nevertheless, these errors can be mitigated by employing a stabilizer state with a larger size $n$, as evidenced by Fig.~\ref{fig:zeeman_bias}(c). For instance, as analyzed in~\ref {appdix:iidp}, when $n = 8$, the probability of QEC misinterpreting syndromes per time unit decreases significantly, ranging from $1.12 \times 10^{-13}$ to $4.31 \times 10^{-8}$. This highlights the flexibility and scalability of our method for practical applications.

\section{Conclusions}

In quantum metrology, noise that is indistinguishable from the signal limits the effectiveness of QEC. Drawing inspiration from the successful integration of QEC with QEM, yet acknowledging the limitations of VSP in the presence of significantly accumulated noise, we introduce a swap test-based method specifically designed to mitigate the indistinguishable noise. The error scaling is systematically analyzed under dephasing and IIDP noise across different scenarios to observe the corresponding behaviors of bias and variance. Though mitigating errors fundamentally requires exponentially increased sampling cost, our analysis also reveals that quantum-enhanced precision in evolution time can be achieved, provided the accumulated noise remains below certain thresholds.

Furthermore, the efficacy of our method compared to the VSP-based method is demonstrated through both single- and multi-parameter estimation tasks across various scenarios. As noise accumulates, the performance of the VSP-based method converges to that of the original noisy case quickly, while our swap test-based method exhibits a much slower increase in parameter estimation errors. The orders-of-magnitude improvement of our method in precision underscores the robust and reliable performance of our method for practical applications.


\appendix

\renewcommand{\thesection}{Supplementary Note \Alph{section}}
\renewcommand{\thesubsection}{\Alph{section}.\arabic{subsection}}

\renewcommand{\figurename}{Supplementary Fig.}
\renewcommand{\thefigure}{\arabic{figure}} 
\setcounter{figure}{0} 

\renewcommand{\tablename}{Supplementary Table}
\renewcommand{\thetable}{\arabic{table}}
\setcounter{table}{0} 

\setcounter{equation}{0} 
\renewcommand{\theequation}{S\arabic{equation}} 

\section{Variance under the IIDP noise}\label{appdix:variance_iidp}

To calculate the variance of the estimator $\hat{\lambda}$, the first-order Taylor expansion around $\langle A\rangle_{\tilde{\rho}}$ and $\tr(\tilde{\rho}^2)$ can be employed to approximate $\hat{\lambda}$. Specifically, we have
\begin{equation}
    \hat{\lambda}\approx \lambda+\left(\frac{\partial \hat{\lambda}}{\partial\hat{A}}\right)\left(\hat{A}-\langle A\rangle_{\tilde{\rho}}\right)+\left(\frac{\partial \hat{\lambda}}{\partial\hat{X}}\right)\left(\hat{X}-\langle X^{(1)}\rangle_{\sigma}\right),
\end{equation}
where $\sigma=\ket{0}\bra{0}\otimes\tilde{\rho}\otimes\tilde{\rho}$. Then, it follows that
\begin{equation}\label{eq:var_lamb}
    {\rm Var}(\hat{\lambda})\approx \left(\frac{\partial \hat{\lambda}}{\partial\hat{A}}\right)^2{\rm Var}(\hat{A}) + \left(\frac{\partial \hat{\lambda}}{\partial \hat{X}}\right)^2{\rm Var}(\hat{X}) + 2\left(\frac{\partial \hat{\lambda}}{\partial\hat{A}}\right)\left(\frac{\partial \hat{\lambda}}{\partial\hat{X}}\right){\rm Cov}(\hat{A}, \hat{X}).
\end{equation}
To evaluate ${\rm Var}(\hat{\lambda})$, it first holds that
\begin{equation}\label{eq:var_sub1}
    {\rm Var}(\hat{A})=\frac{1}{2\nu}(\langle A^2\rangle_{\tilde{\rho}}-\langle A\rangle_{\tilde{\rho}}^2)\le\frac{1}{2\nu}\Vert A\Vert_{\infty}=\frac{1}{2\nu}
\end{equation}
and 
\begin{equation}\label{eq:var_sub2}
    {\rm Var}(\hat{X})=\frac{1}{2}(\langle I\rangle_{\sigma}-\langle X^{(1)}\rangle_{\sigma}^2)=1-\tr(\tilde{\rho}^2)^2\le \frac{4-p''^4}{4\nu}\le\frac{1}{\nu},
\end{equation}
In addition, note that 
\begin{equation}
\begin{aligned}
    \langle X^{(1)}A^{(2)}\rangle_{\sigma}&=\langle A\rangle_{\tilde{\rho}^2}\\
    &=p''^2\langle A\rangle_{\tilde{\rho}_z^2}+2p''(1-p'')\langle A\rangle_{\tilde{\rho}_z\tilde{\rho}_x}+(1-p'')^2\langle A\rangle_{\tilde{\rho}_x^2}\\
    &\le p''^2\langle A\rangle_{\tilde{\rho}_z^2}+2p''(1-p'')+(1-p'')^2
\end{aligned}
\end{equation}
and 
\begin{equation}\label{eq:var_sub6}
\begin{aligned}
    \langle X^{(1)}\rangle_{\sigma}\langle A^{(2)}\rangle_{\sigma}&=\tr(\tilde{\rho}^2)\langle A\rangle_{\tilde{\rho}}\\
    &=\left(p''^2\langle X^{(1)}\rangle+2p''(1-p'')\tr(\tilde{\rho}_z\tilde{\rho}_x)+(1-p'')^2\tr(\tilde{\rho}_x^2)\right)\\
    &\quad\times\left(p''\langle A\rangle_{\tilde{\rho}_z}+(1-p'')\langle A\rangle_{\tilde{\rho}_x}\right)\\
    &\ge p''^3\langle X^{(1)}\rangle\langle A\rangle_{\tilde{\rho}_z}-p''^2(1-p'')\langle X^{(1)}\rangle
\end{aligned},
\end{equation}
where we assume a large $p''$ such that $\langle A\rangle_{\tilde{\rho}}\ge 0$. Therefore, the covariance between $\hat{A}$ and $\hat{X}$ is upper bounded by
\begin{equation}\label{eq:var_sub3}
\begin{aligned}
    {\rm Cov}(\hat{A},\hat{X})=&\frac{1}{\nu}\left(\langle X^{(1)}A^{(2)}\rangle_{\sigma}-\langle X^{(1)}\rangle_{\sigma}\langle A^{(2)}\rangle_{\sigma}\right)\\
    \le& \frac{1}{\nu}\left(p''^2\langle A\rangle_{\tilde{\rho}_z^2}+2p''(1-p'')+(1-p'')^2\right.\\
    &\left.-p''^3\langle X^{(1)}\rangle\langle A\rangle_{\tilde{\rho}_z}+p''^2(1-p'')\langle X^{(1)}\rangle\right)\\
    =&\frac{1}{\nu}\left(p''^2\langle A\rangle_{\tilde{\rho}_z}\left(1-p''\langle X^{(1)}\rangle\right)+(1-p'')\left(1+p''+p''^2\langle X^{(1)}\rangle\right)\right)
\end{aligned}.
\end{equation}
According to Eq.~\eqref{eq:var_lamb}, we have
\begin{equation}\label{eq:var_biased_lamb}
\begin{aligned}
    {\rm Var}(\hat{\lambda})\approx&\left(4n^2t^2\left(2\langle X^{(1)}\rangle_{\sigma}-1-\langle A\rangle_{\tilde{\rho}}^2\right)\right)^{-1}{\rm Var}(\hat{A})\\
    &+\langle A\rangle_{\tilde{\rho}}^2\left(4n^2t^2\left(2\langle X^{(1)}\rangle_{\sigma}-1\right)^2\left(2\langle X^{(1)}\rangle_{\sigma}-1-\langle A\rangle_{\tilde{\rho}}^2\right)\right)^{-1}{\rm Var}(\hat{X})\\
    &+\langle A\rangle_{\tilde{\rho}}\left(2n^2t^2\left(2\langle X^{(1)}\rangle_{\sigma}-1\right)\left(2\langle X^{(1)}\rangle_{\sigma}-1-\langle A\rangle_{\tilde{\rho}}^2\right)\right)^{-1}{\rm Cov}(\hat{A},\hat{X})\\
    \le&\left(2\nu n^2t^2\left(2\langle X^{(1)}\rangle_{\sigma}-1-\langle A\rangle_{\tilde{\rho}}^2\right)\right)^{-1}\left(\frac{1}{4}+\frac{\langle A\rangle_{\tilde{\rho}}^2}{2\left(2\langle X^{(1)}\rangle_{\sigma}-1\right)^2}+\frac{\langle A\rangle_{\tilde{\rho}}{\rm Cov}(\hat{A},\hat{X})}{2\langle X^{(1)}\rangle_{\sigma}-1}\right)
\end{aligned}.
\end{equation}

To further evaluate the variance of $\hat{\lambda}$, the involved terms $2\langle X^{(1)}\rangle_{\sigma}-1$ and $\langle A\rangle_{\tilde{\rho}}$ can be bounded by
\begin{equation}\label{eq:var_sub4}
\begin{aligned}
    2\langle X^{(1)}\rangle_{\sigma}-1=&2\tr(\tilde{\rho}^2)-1\\
    \ge&2p''^2\langle X^{(1)}\rangle-1\\
    =&p''^2(1-2p_{\rm odd})^2-(1-p''^2)
\end{aligned}
\end{equation}
and
\begin{equation}\label{eq:var_sub5}
\begin{aligned}
    \langle A\rangle_{\tilde{\rho}}=&p''\langle A\rangle_{\tilde{\rho}_z}+(1-p'')\langle A\rangle_{\tilde{\rho}_x}\\
    \le&p''\langle A\rangle_{\tilde{\rho}_z}+(1-p'')
\end{aligned},
\end{equation}
respectively. Substituting Eqs.~\eqref{eq:var_sub3},~\eqref{eq:var_sub4}, and~\eqref{eq:var_sub5} into Eq.~\eqref{eq:var_biased_lamb}, 
the upper bound of ${\rm Var(\hat{\lambda})}$ is obtained.

\section{Multi-parameter estimation}\label{appdix:multiparas}

For the probe state $\ket{\psi_0}=\frac{1}{\sqrt{2}}(\ket{0}^{\otimes n}+\ket{1}^{\otimes n})$, the final state evolved after the encoding unitary $U_{\blamb}=e^{-itH_{\blamb}}$, where $H_{\blamb}=\lambda_1X^{(1)}X^{(2)}\cdots X^{(n)}+\lambda_2\sum_{i=1}^nZ^{(i)}$, reads
\begin{equation}
    \ket{\psi}=\frac{1}{\sqrt{2}}\left(\left(\cos(\Omega t)-i\frac{\lambda_1+n\lambda_2}{\Omega}\sin(\Omega t)\right)\ket{0}^{\otimes n}+\left(\cos(\Omega t)-i\frac{\lambda_1-n\lambda_2}{\Omega}\sin(\Omega t)\right)\ket{1}^{\otimes n}\right),
\end{equation}
where $\Omega=\sqrt{\lambda_1^2+n^2\lambda_2^2}$. Consider the dephasing channel $\mathcal{E}_z^{(i)}(\rho)=(1-p_z)\rho+p_zZ^{(i)}\rho Z^{(i)}$ and define $\ket{\phi}=Z^{(i)}\ket{\psi}$, it follows that
\begin{equation}
\begin{aligned}
    \vert\langle\psi|\phi\rangle\vert^2=&\frac{1}{4}\left|\frac{(\lambda_1+n\lambda_2)^2}{\Omega^2}\sin^2(\Omega t)-\frac{(\lambda_1-n\lambda_2)^2}{\Omega^2}\sin^2(\Omega t)\right|^2\\
    =&4n^2\lambda_1^2\lambda_2^2\sin^4(\Omega t)/\Omega^4
\end{aligned}.
\end{equation}
When $\Omega t$ is small, we have $\vert\langle\psi|\phi\rangle\vert^2 \approx 16n^2\lambda_1^2\lambda_2^2t^4$, due to the approximation $\sin(\Omega t) \approx \Omega t$. Consequently, for local estimation tasks with small $\lambda_i$, the noisy state $\ket{\phi}$ is nearly orthogonal to the noiseless state $\ket{\psi}$. Thus, the probability $p_{\rm odd}$ can still be estimated using the formula presented in the main text with small bias.

Moreover, when this condition does not hold, the utilization of the expectation value of the swap test can be adjusted to achieve an unbiased estimation of $\hat{\blamb}$. Specifically, we have
\begin{equation}
\begin{aligned}
    \tilde{\rho}_z&=\circ_i\mathcal{E}_z^{(i)}(\ket{\psi}\bra{\psi})=(1-p_{\rm odd})\ket{\psi}\bra{\psi}+p_{\rm odd}\ket{\phi}\bra{\phi}
\end{aligned},
\end{equation}
where $p_{\rm odd}=\mathlarger{\sum}_{i{\rm\ is\ odd}}\begin{pmatrix}
    n \\
    i
\end{pmatrix}p_z^i(1-p_z)^{n-i}=\left(1-(1-2p_z)^n\right)/2$.
Consequently, the expectation value of the swap test can be expressed as
\begin{equation}\label{eq_sup:x_exp}
\begin{aligned}
    \langle X^{(1)}\rangle=\tr(\tilde{\rho}_z^2)=2(1-\alpha)p_{\rm odd}^2+2(\alpha-1)p_{\rm odd}+1
\end{aligned},
\end{equation}
where $\alpha=\vert\langle\psi|\phi\rangle\vert^2$. Thus, assume $p_z<1/2$, it can be easily verified that
\begin{equation}\label{eq_sup:p_odd}
    p_{\rm odd}=\frac{1-\alpha-\sqrt{(1-\alpha)(2\langle X^{(1)}\rangle-1-\alpha})}{2(1-\alpha)}.
\end{equation}
Since $\alpha$ contains unknown parameters $\blamb$, $p_{\rm odd}$ can not be directly estimated using the swap test. However, it can be parameterized by $\blamb$ and $\langle X^{(1)} \rangle$ based on Eq.~\eqref{eq_sup:p_odd}. By applying maximum likelihood estimation to this parameterization and the associated state evolution, we can obtain an unbiased estimator of $\blamb$.

\begin{figure}[t]
  \centering
  \includegraphics[width=\columnwidth]{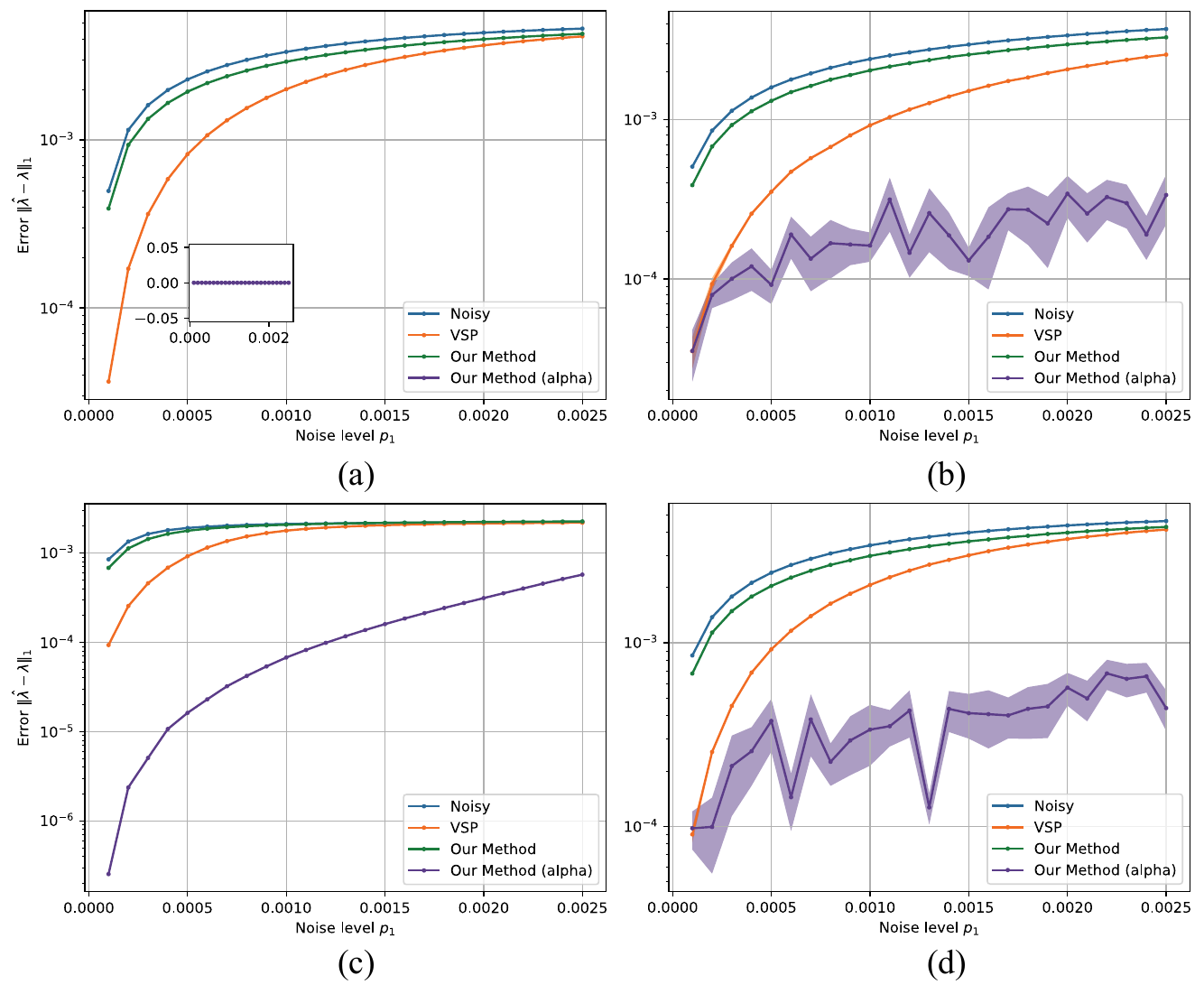}
  \caption{\textbf{Multi-parameter estimation errors for different methods under large values of $\lambda_i$.} The absolute parameter estimation errors, $\Vert\hat{\blamb} - \blamb\Vert_1$, are shown for the original noisy (blue), VSP-based (orange), our methods based on Eq.~\eqref{eq:tr2_qec} (green) and Eq.~\eqref{eq_sup:x_exp} (purple). Under dephasing noise, panel (a) shows results with infinite shots, and panel (b) shows results with $10^6$ shots. Under IIDP noise, panel (c) shows results with infinite shots, and panel (d) shows results with $10^6$ shots. For each scenario, numerical simulations were performed 10 times to determine the mean absolute errors, indicated by solid lines, and their corresponding 95\% confidence intervals, shown as shaded regions.}\label{fig_sup:multi_alpha}
\end{figure}

As an example, we set $\blamb=(1,\ 0.25)$, and repeat the numerical simulations described in Sec.~\ref{subsec:multi} to observe the performance of our modified method based on Eq.~\eqref{eq_sup:x_exp}. Supplementary Figs.~\ref{fig_sup:multi_alpha}(a) and (b) illustrate the corresponding results for different methods under dephasing noise with infinite and $10^6$ shots, respectively. In this case, since $\alpha$ deviates significantly from zero, the direct application of our method based on Eq. \eqref{eq:tr2_qec} yields inferior results compared to the VSP-based method, as indicated by the green lines. Conversely, by employing our method based on Eq. \eqref{eq_sup:x_exp}, we achieve an unbiased estimate of $\blamb$ in the infinite-shot limit and a substantially reduced estimation error with finite shots, as evidenced by the purple lines.

In addition, when IIDP noise is considered, the corresponding results are plotted in Supplementary Figs.~\ref{fig_sup:multi_alpha}(c) and (d). Again, the errors increase, compared to the dephasing noise cases, because QEC would misinterpret syndromes, leading to an accumulation of bit-flip errors. Nevertheless, our modified method demonstrates a significant reduction in bias across all cases. 

\section{Analysis on quantum error corrected IIDP noise}\label{appdix:iidp}

Recall that the independent, identically distributed Pauli (IIDP) noise is expressed as
\begin{equation}\label{supeq:iidp}
    \mathcal{E}^{(i)}(\rho)=p_I\rho+p_xX^{(i)}\rho X^{(i)}+p_yY^{(i)}\rho Y^{(i)}+p_zZ^{(i)}\rho Z^{(i)},
\end{equation}
and denote $\mathcal{F}_{\rm QEC}$ as the quantum operator applying quantum error correction (QEC). Specifically, define $P_{\sigma}^{(S_{\sigma})}=\otimes_{i\in S_{\sigma}}P_{\sigma}^{(i)}$, where $\sigma\in\{I,X,Y,Z\}$. Then, for an $n$-qubit probe state $\ket{\psi}\in{\rm span}\{\ket{0}^{\otimes n},\ket{1}^{\otimes n}\}$ evolving through ${\mathcal{E}^{(i)}}^{\otimes n}$, it could results in a noisy state $ Z^{(S_Z)}Y^{(S_Y)}X^{(S_X)}\rho X^{(S_X)}Y^{(S_Y)}Z^{(S_Z)}$. If $|S_X|+|S_Y|<\lceil n/2\rceil$, $\mathcal{F}_{\rm QEC}$ can detect the positions $X^{(i)}$ and $Y^{(i)}$ errors happen, and apply the operator $X^{(S_X\cup S_Y)}$ to correct errors, resulting in the error-corrected state $ Z^{(S_Y\cup S_Z)}\rho Z^{(S_Y\cup S_Z)}$. Otherwise, $\mathcal{F}_{\rm QEC}$ would misinterpret errors and apply the operator $X^{(S_I\cup S_Z)}$ on the noisy state, leading to the state $ Z^{(S_Y\cup S_Z)}X^{\otimes n}\rho X^{\otimes n} Z^{(S_Y\cup S_Z)}$. 

To fully characterize $\tilde{\rho}=\mathcal{F}_{\rm QEC}\circ {\mathcal{E}^{(i)}}^{\otimes n}(\ket{\psi}\bra{\psi})$, we first consider the probability that the noisy state admits the form of $ Z^{(S_Y\cup S_Z)}\ket{\psi}\bra{\psi} Z^{(S_Y\cup S_Z)}$ with $|S_Y\cup S_Z|$ being an odd number. Concretely, let $N_{\sigma}$ be the qubit number of the error $\sigma$ happening on. Define $A=N_X+X_Y$ and $B=N_Y+N_Z$, it is equivalently to calculate $P(A<\lceil n/2\rceil,B {\ \rm odd})$. Here, we can track both $A$ and $B$ with a bivariate generating function. For $n=1$, the generating function is
\begin{equation}
\begin{aligned}
    F(x,y)&=p_I\cdot x^0y^0+p_x\cdot x^1y^0+p_y\cdot x^1y^1+p_z\cdot x^0y^1\\
    &=p_I+p_xx+p_yxy+p_zy
\end{aligned},
\end{equation}
where $x$ and $y$ mark increments to $A$ and $B$, respectively. Then, for the $n$-qubit case, we have
\begin{equation}
    G(x,y)=F(x,y)^n=\sum_{a,b=0}^n P(A=a,B=b)x^ay^b.
\end{equation}
Denote $[x^ay^b]G(x,y)$ as the coefficient of the $x^ay^b$ term in $G(x,y)$, it holds that
\begin{equation}
    P(A<\lceil n/2\rceil,B {\ \rm odd})=\sum_{a=0}^{\lceil n/2\rceil-1}\sum_{b\ { \rm odd}}[x^ay^b]G(x,y).
\end{equation}
For a given $a$, it can be easily verified that
\begin{equation}
    \sum_{b\ { \rm odd}}[x^ay^b]G(x,y)=\frac{1}{2}\left([x^a]G(x,1)-[x^a]G(x,-1)\right).
\end{equation}
Therefore, we have
\begin{equation}
\begin{aligned}
    &P(A<\lceil n/2\rceil,B {\ \rm odd})\\=&\frac{1}{2}\left(\sum_{a=0}^{\lceil n/2\rceil-1}[x^a]G(x,1)-\sum_{a=0}^{\lceil n/2\rceil-1}[x^a]G(x,-1)\right)\\
    =&\frac{1}{2}\left(\sum_{a=0}^{\lceil n/2\rceil-1}\begin{pmatrix} n\\a \end{pmatrix}(p_X+p_Y)^a(p_I+p_Z)^{n-a}-\sum_{a=0}^{\lceil n/2\rceil-1}\begin{pmatrix} n\\a \end{pmatrix}(p_X-p_Y)^a(p_I-p_Z)^{n-a}\right)
\end{aligned}.
\end{equation}
Besides, the probability of a successful QEC is $P(A<\lceil n/2\rceil)=\sum_{a=0}^{\lceil n/2\rceil-1}\begin{pmatrix} n\\a \end{pmatrix}(p_X+p_Y)^a(p_I+p_Z)^{n-a}$, and it holds that $P(A<\lceil n/2\rceil,B {\ \rm even})=P(A<\lceil n/2\rceil)-P(A<\lceil n/2\rceil,B {\ \rm odd})$. The probabilities $P(A\ge \lceil n/2\rceil,B {\ \rm odd})$ and $P(A\ge \lceil n/2\rceil,B {\ \rm even})$ can be derived analogously. 

To simplify the expression, we redefine $p_{x}'=P(A\ge\lceil n/2\rceil),\ p_{{\rm odd}|x}=P(B {\ \rm odd}|A\ge\lceil n/2\rceil)$ and $p_{{\rm odd}|z}=P(B {\ \rm odd}|A<\lceil n/2\rceil)$. Then, the noisy state reads
\begin{equation}
\begin{aligned}
    \tilde{\rho}=&\mathcal{F}_{\rm QEC}\circ {\mathcal{E}^{(i)}}^{\otimes n}(\ket{\psi}\bra{\psi})\\
    =&(1-p'_x)\left((1-p_{{\rm odd}|z})\ket{\psi}\bra{\psi}+p_{{\rm odd}|z}Z^{(1)}\ket{\psi}\bra{\psi}Z^{(1)}\right)\\&+p'_x\left((1-p_{{\rm odd}|x})X^{\otimes n}\ket{\psi}\bra{\psi}X^{\otimes n}+p_{{\rm odd}|x}Z^{(1)}X^{\otimes n}\ket{\psi}\bra{\psi}X^{\otimes n}Z^{(1)}\right)
\end{aligned}.
\end{equation}

\bibliography{ref}

\end{document}